# Measuring institutional research productivity for the life sciences: the importance of accounting for the order of authors in the byline[1]

*Giovanni Abramo*[a,b,*], *Ciriaco Andrea D'Angelo*[b], *Francesco Rosati*[b]

[a] Institute for System Analysis and Computer Science (IASI-CNR)
National Research Council of Italy

[b] Laboratory for Studies of Research and Technology Transfer
School of Engineering, Department of Management
University of Rome "Tor Vergata"

**Abstract**

Accurate measurement of institutional research productivity should account for the real contribution of the research staff to the output produced in collaboration with other organizations. In the framework of bibliometric measurement, this implies accounting for both the number of co-authors and each individual's real contribution to scientific publications. Common practice in the life sciences is to indicate such contribution through the order of author names in the byline. In this work, we measure the distortion introduced to university-level bibliometric productivity rankings when the number of co-authors or their position in the byline is ignored. The field of observation consists of all Italian universities active in the life sciences (Biology and Medicine). The analysis is based on the research output of the university staff over the period 2004-2008. Based on the results, we recommend against the use of bibliometric indicators that ignore co-authorship and real contribution of each author to research outputs.

**Keywords**

*Research evaluation; co-authorship; fractional counting; bibliometrics; biology; medicine*

* **Corresponding author:** Dipartimento di Ingegneria dell'Impresa, Università degli Studi di Roma "Tor Vergata", Via del Politecnico 1, 00133 Rome - ITALY, tel/fax +39 06 72597362, giovanni.abramo@uniroma2.it

# 1. Introduction

Research managers and policy makers are constantly confronted with strategic and operational choices, particularly in the area of allocative efficiency for their various national systems. There is currently a broad trend to draw on bibliometric techniques for support in such decision-making, thus the delicate and related task of evaluating production efficiency is also taking on growing importance.

The main indicator of efficiency for almost any activity is productivity, or in very simple terms, the ratio between the value of output produced and the value of inputs needed to produce it. As for any measurement system, that for research productivity is subject to limits and approximations. The outputs tend to be of an intangible nature, and a further concern is that they are generally obtained through collaboration of various individuals and institutions. In evaluating the scientific activity of an organization it is thus fundamental to identify the real contribution of the research staff to the outputs, including those produced in collaboration with other organizations. In the scientific fields where codification of results is primarily through publication in scientific journals, indexed in such databases as Web of Science (WoS) or Scopus, bibliometrics can be conveniently applied for large scale evaluation of productivity. In this case, the contribution of scientists and organizations to the outputs of research projects can be recognized through the analysis of co-authorship of publications. In the life sciences in particular, widespread practice is for the authors to signal their various contributions to the results of the published research by the positioning of the names in the byline.

In this work, we propose to measure the distortions encountered in the evaluation of organizational research productivity in the life sciences when no consideration is given to the number of co-authors of the research works and their order in the byline. As much as it may seem logical and even mandatory under economic theory of production that one would account for both factors in comparative measurement of research productivity, it is actually not at all rare that they are partially or completely ignored. In national research evaluation exercises with peer-review techniques, this is standard practice: for example in the UK research assessment exercise series (RAEs) and in the Italian triennial evaluation of research (VTR), peer evaluators are only called to judge the level of excellence of the products that the researchers submit, independent of true extent of the author's contribution to their accomplishment. The same is true for some national exercises conducted with bibliometric techniques, such as the current VQR in Italy. Even famous and widely used bibliometric performance indicators, like the $h$-index (Hirsch, 2005) and the $g$-index (Egghe, 2006), totally ignore any consideration of the contributions of the individual authors to the scientific product. Little attention has been paid to advice from the index "inventors", such as Hirsch (2005), who warned that "subfields with typically large collaborations (e.g., high-energy experiment) will exhibit larger $h$ values", and further recommended that "in cases of large differences in the number of co-authors, it may be useful in comparing different individuals to normalize $h$ by a factor that reflects the average number of co-authors". Little attention has also been paid to the specific corrections proposed, such as the simple division of the $h$-index by the average number of co-authors included in the Hirsch core (Batista et al., 2006; Egghe, 2008), or consideration of the actual number of co-authors and the scientists' relative position in the byline (Wan et al., 2008).

There is clearly growing agreement among bibliometricians on the desirability of accounting for co-authorship through fractional counting (Aksnes et al., 2012; He et al., 2012; Carbone, 2011), though there are still differences over the most appropriate fraction to assign to each co-author (Gauffriau et al., 2008; Laurance, 2006; Verhagen et al., 2003; Bhandari et al., 2003).

This work is not precisely concerned with establishing the most appropriate credit to assign to contributions from co-authors. Rather after choosing fixed, but potentially "fine-tunable", criteria to assign different weight to the various positions in the byline, the objective we set is to measure the extent of the distortion in performance ranking when the number of co-authors and their order are totally ignored. In Italy there are no fixed guide-lines establishing the order of names in the byline, even though some important academic committees have officially pronounced themselves in favor [2], particularly concerning life sciences. The Italian National University Council states that the biological and medical sciences are characterized by "scientific works that are prevailingly by multiple authors, in which the first and last authors are generally the leader of the specific research and the leader of the entire research group, and where in certain fields the second name indicates the co-leader of the specific research". In effect, wide-spread practice is that the position of first author falls to the "idea generator" and person who executes the bulk of the work, while the last position is assigned to the overall working-group leader. In the case of multiple authors from more than one institution, with similar contributions to the research, the indication of second and second-last authors also becomes significant. In general, if the position of the first author is assigned to one organization, the last name listed will be that of the group leader from the other institution, and the positions of second and second-last authors are then assigned in the opposite manner.

In a recent work, applying the criteria just described, Abramo et al. (2012) measured the distortion introduced in individual productivity rankings when the number of co-authors or their position in the byline is ignored. In the current work we wish to determine the extent of distortion at a higher level of analysis: that of the organization. Given the compensatory effects of aggregation we would expect the distortion to be reduced.

The field of observation is based on 2004-2008 scientific production by professors in Biology and Medicine from all Italian universities. We will calculate the universities research productivity and draw up a total of six ranking lists, three for each of two types of bibliometric indicators, based on number of publications and number of citations: i) a list that takes account of both number of co-authors of each publication and their position in the list; ii) a list that does not consider position; iii) one that does not consider co-authorship in any way.

The next section illustrates the methodology and dataset used for the analyses. Section 3 presents the results concerning the correlations between the ranking lists, the analysis of shifts in position in the rankings both at the field and at discipline levels, and a deeper examination concerning the upper quartile of universities active in each of Medicine and Biology. The work concludes with a summary of the results and the authors' considerations.

---

[2] http://www.cun.it/media/100033/area6.pdf, last accessed on Jan. 23, 2013.

## 2. Methodology

### 2.1 Measuring research productivity

Research activity is a production process in which the inputs consist of human and tangible resources (scientific instruments, materials, etc.) and intangible resources (accumulated knowledge, social networks, etc.), and where outputs have a complex character of both tangible nature (publications, patents, conference presentations, databases, protocols, etc.) and intangible nature (tacit knowledge, consulting activity, etc.). The new-knowledge production function therefore has a multi-input and multi-output character. The principal efficiency indicator of any production system is labor productivity. To calculate labor productivity in this context we need to adopt some simplifications and assumptions. In the hard sciences, including life sciences, the prevalent form of codification of research output is publication in scientific journals. As a proxy of total output, in this work we consider only publications (articles, article reviews and proceeding papers) indexed in the WoS; as proxy of value of this output we consider citations received. The other forms of output which we neglect are often followed by publications that describe their content in the scientific arena, so the analysis of publications alone in many cases actually avoids a potential double counting.

When measuring labor productivity, if there are differences in the resources available to each scientist then one should normalize by them. Unfortunately, relevant data are not available at individual level in Italy. Here we assume that resources available to professors within the same field of observation are the same. The second assumption is that the hours devoted to research are more or less the same for all professors. In Italy the above assumptions are acceptable because in the period of observation, core government funding was input oriented and distributed to satisfy the resource needs of each and every university in function of their size and activities. Furthermore, the hours that each professor has to devote to teaching are established by national regulations and are the same for all.

As noted above, research projects frequently involve a team of researchers, which shows in co-authorship of publications. Productivity measures then need to account for the fractional contributions of scientists to their outputs. In the life sciences, the order of co-authors in the byline reflects the relative contribution to the project and needs to be weighted accordingly. Furthermore, because the intensity of publications varies across fields (Abramo et al., 2008), in order to avoid distortions in productivity rankings, we compare researchers within the same field. A prerequisite of any distortion-free research performance assessment is thus a classification of each researcher in one and only one field, a practice that happens to be a feature (perhaps unique in the world) of the Italian higher education system. In the Italian system the hard sciences are represented by 205 such fields (named scientific disciplinary sectors, SDSs[3]), grouped into nine disciplines (named university disciplinary areas, UDAs[4]). The life sciences in particular are grouped under two UDAs, Medicine and Biology, consisting respectively of 50 and 19 SDSs. In the next section we propose the productivity indicators for ranking the universities active in each field and discipline of the life sciences.

---

[3] The complete list is accessible at http://attiministeriali.miur.it/UserFiles/115.htm, Last accessed on Jan. 23, 2013.
[4] Mathematics and computer sciences; physics; chemistry; earth sciences; biology; medicine; agricultural and veterinary sciences; civil engineering; industrial and information engineering.

## 2.2 Indicators

### 2.2.1 Productivity at the SDS level

We adopt university productivity measures of two types: a gross one based on publication counts, named "weighted fractional output", or WFO; and a more sophisticated and accurate one based on field-normalized citations, named "weighted fractional impact", WFI.

In formulae, the WFO of a generic university in the SDS *s* is:

$$WFO_s = \frac{1}{RS_s} \cdot \sum_{i=1}^{N_s} wf_{i,s}$$

[1]

Where:

$wf_{i,s}$ = weighted fractional contribution of researchers in SDS *s* of the university, as co-authors of publication *i*. Different contributions are given to each co-author according to his/her position in the byline and the character of the co-authorship (intra-mural or extra-mural). If first and last authors belong to the same university, 40% of the publication is attributed to each of them; the remaining 20% is divided among all other co-authors. If the first two and last two authors belong to different universities, 30% of the publication is attributed to first and last authors; 15% of the publication is attributed to second and second-last author; the remaining 10% is divided among all others[5].

$N_s$ = number of publications of the research staff in SDS *s* of the university, in the period of observation

$RS_s$ = research staff of the university in SDS *s*, in the observed period.

The WFI is:

$$WFI_s = \frac{1}{RS_s} \cdot \sum_{i=1}^{N_s} \frac{c_i}{Me_i} * wf_{i,s}$$

[2]

Where:

$c_i$ = citations received by publication *i* (observed at June 30, 2009);

$Me_i$ = median[6] of the distribution of citations received for all Italian cited publications of the same year and subject category of publication *i*;

$wf_{i,s}$; $Ns$ and $RS_s$ = same as in [1].

Based on the above indicators, we measure a further four: two of these, FO and FI, eliminate the weighting that takes account of the position in the byline of co-authors and consider a fractional contribution invariably equal to the reciprocal of the number of co-authors; and two, O and I, eliminate the fractional count that takes account of co-authors. For each indicator, we then elaborate university ranking lists for each SDS. To compare productivity in different SDSs, we express university productivity on a percentile scale of 0-100 (worst to best) for comparison of absolute values of indicators calculated for all Italian universities active in the same SDS.

---

[5] The weighting values for both this indicator and the WFI indicator below were assigned based on the results of interviews with top Italian professors in the life sciences. The values could be changed to suit different practices in other national contexts.

[6] We standardize citations by the median, since as frequently observed in literature (Lundberg, 2007), standardization of citations with respect to median value rather than to the average is justified by the fact that distribution of citations is highly skewed in almost all disciplines. However we note that there is not yet agreement among bibliometricians on the most efficient scaling factor.

**2.2.2 Productivity at the UDA level**

At this level of analysis we aggregate productivity measures of the SDSs in each UDA, standardizing them to national average and weighting for the relative size of the SDS. In this way, we take account of the varying intensity of publication and citation for the SDSs, avoiding the typical distortion of measures at the aggregate level (Abramo et al., 2008).

In formulae, the total weighted fractional output of a general university ($WFO_u$) in the UDA $u$ (Biology or Medicine) is then:

$$WFO_u = \sum_{s=1}^{N_u} \frac{WFO_s}{\overline{WFO_s}} \cdot \frac{RS_s}{RS_u}$$

[3]

With:
$RS_u$ = research staff of the university in the UDA $u$, in the observed period;
$N_u$ = number of SDSs of the university in the UDA $u$;
$\overline{WFO_s}$ = average value of the weighted fractional output of all Italian universities in SDS $s$.
$WFO_s$ and $RS_s$ = same as in [1]

At the same time, the total weighted fractional impact of a general university ($WFI_u$) in the UDA $u$ (Biology or Medicine) is:

$$WFI_u = \sum_{s=1}^{N_u} \frac{WFI_s}{\overline{WFI_s}} \cdot \frac{RS_s}{RS_u}$$

[4]

With:
$\overline{WFI_s}$ = average value of the weighted fractional impact of all Italian universities in SDS $s$.
$WFI_s$ and $RS_s$ = same as in [2]
$RS_u$ and $N_u$ = same as in [3]

In the case of the UDAs we again measure a further four indicators: two eliminating the weighting that takes account of the order in the byline of co-authors; and two eliminating the fractional count that takes account of co-authors. For each indicator, we then elaborate university ranking lists in Biology and Medicine expressed on a percentile scale of 0-100 (worst to best) for comparison of absolute values of indicators calculated for all Italian universities active in the same UDA.

**2.3 Datasets**

Data on the research staff of each institution and their assignment to SDSs are extracted from the database on Italian university personnel, maintained by the Ministry for Universities and Research[7]. The bibliometric dataset used to measure productivity is extracted from the Italian Observatory of Public Research (ORP)[8], a database developed and maintained by the authors and derived under license from the Thomson Reuters WoS. Beginning from the raw data of the WoS and applying a complex algorithm for disambiguation of the true identity of the authors and their institutional affiliations, each publication is attributed to the university scientist or scientists that produced it (D'Angelo et al., 2011).

Our overall dataset includes all the universities active in those Biology and Medicine SDSs (Appendix A) where bibliometric techniques provide a robust calculation of productivity[9]: there are a total of 19 such SDSs in Biology and 46 in Medicine.

From this, we now prepare two data subsets necessary to conduct the analyses. The first includes all and only the universities with non-nil output (in WFO, FO, O), meaning those where the research staff produced at least one publication over the period 2004-2008. The second set includes all and only the universities with non-nil impact (WFI, FI, I), meaning those where staff produced at least one cited publication (as of 30/06/2009) over the 2004-2008 period. For the purposes of our analyses it would not make sense to consider any universities with nil output or citations, given that this means nil productivity, regardless of the counting method considered. Overall, the WFO dataset thus counts 64 universities while the WFI dataset (fully contained within the WFO dataset) includes 62 universities (Table 1).

*Table 1: Datasets for the analyses*

| UDA | n. of SDSs | Research staff[*] | Publications[**] | Citations | Universities | |
|---|---|---|---|---|---|---|
| | | | | | WFO | WFI |
| Biology | 19 | 4,718 | 27,600 | 218,105 | 63 | 61 |
| Medicine | 46 | 8,940 | 50,331 | 407,311 | 54 | 54 |
| Total | 65 | 13,658 | 70,740[***] | 563,201[**] | 64 | 62 |

[*] *Counting only staff with at least one publication in the 2004-2008 period*
[**] *Number of publications indicated as under authorship of at least one researcher in the UDA*
[***] *Total values differ from the column sums due to the effect of multiple counts for publications co-authored by researchers in both Biology and Medicine.*

**3. Results**

In this section we will contrast the three ranking lists derived from each type of productivity indicator: those based on simple publication count (WFO, FO, O) and those based on standardized citations (WFI, FI, I). First we present the case of a single SDS, then extend the analysis to all SDSs, and finally to the data for the aggregate UDAs. We begin with the correlation analyses between the ranking lists, then continue with the analyses of shifts in position when changing from ranking under one indicator to rankings under another, and conclude with a deeper analysis of the top 25% of universities.

---
[7] http://cercauniversita.cineca.it/php5/docenti/cerca.php. Last accessed on Jan. 23, 2013.
[8] www.orp.researchvalue.it. Last accessed on Jan. 23, 2013.
[9] To ensure the representativity of publications as proxy of the research output, the analysis excludes those SDSs (MED/02; MED/43 and MED/47) where less than 50% of researchers produced a WoS-indexed publication over the period 2004-2008. Further, we exclude MED/48 since the research staff of this SDS is distributed in only seven universities.

### 3.1 Correlation analysis and quartile variations in productivity rankings for SDS

In this section we apply formulas [1] and [2] as presented in part 2.1 to calculate the indicators WFI, FI and I, and then compare the SDS ranking lists derived from the indicators.

As an example we present the case of BIO/12 (Clinical biochemistry and molecular biology), in the UDA Biology (Table 2). Columns 2, 3 and 4 show the absolute values of WFI, FI and I (with the ranks in brackets), for each of the 38 universities with research staff in this SDS during the 2004-2008 period. The correlation index between WFI and I is 0.885, slightly lower than that between WFI and FI, at 0.900. We also observe an almost perfect correlation between FI and I, with the correlation index at 0.940.

*Table 2: Values of WFI, FI and I for universities active in BIO/12 (Clinical biochemistry and biology), ordered by decreasing WFI (ranks for FI and I in brackets)*

| University | WFI | FI | I |
|---|---|---|---|
| ID1 | 4.82 | 1.54 (3) | 9.96 (2) |
| ID2 | 4.68 | 3.02 (1) | 11.76 (1) |
| ID3 | 3.20 | 1.56 (2) | 5.54 (7) |
| ID4 | 2.01 | 1.20 (4) | 7.60 (4) |
| ID5 | 1.78 | 0.46 (10) | 3.34 (9) |
| … | … | … | … |
| ID34 | 0.14 | 0.06 (33) | 0.28 (33) |
| ID35 | 0.12 | 0.03 (35) | 0.17 (35) |
| ID36 | 0.12 | 0.01 (37) | 0.13 (36) |
| ID37 | 0.04 | 0.02 (36) | 0.11 (37) |
| ID38 | 0.03 | 0.01 (38) | 0.05 (38) |

We repeat the same analysis for all 65 SDSs and in Figure 1 provide the distribution of Spearman correlation indexes for rankings based on WFI, FI and I. We observe that all the SDSs show correlations greater than 0.8.

Examining the WFI to I comparison, there are 9 SDSs (13.8% of total) with correlation greater than 0.95 and 44 SDSs (67.7%) with correlation greater than 0.9. Further, comparing ranking lists by WFI and FI, 13 SDSs (20.0% of total) show correlation greater than 0.95, and a full 50 SDSs (76.9% of total) show correlation higher than 0.9. Comparing FI and I, there are 34 SDSs (52.3%) which show correlation greater than 0.95, while for 59 SDSs (90.8%) correlation is higher than 0.9. The lowest correlation for the WFI to I comparison (0.829) is seen in MED/37 (Neuroradiology), for WFI to FI comparison (0.826) in BIO/17 (Histology), and for FI to I comparison (0.818) is in BIO/08 (Anthropology).

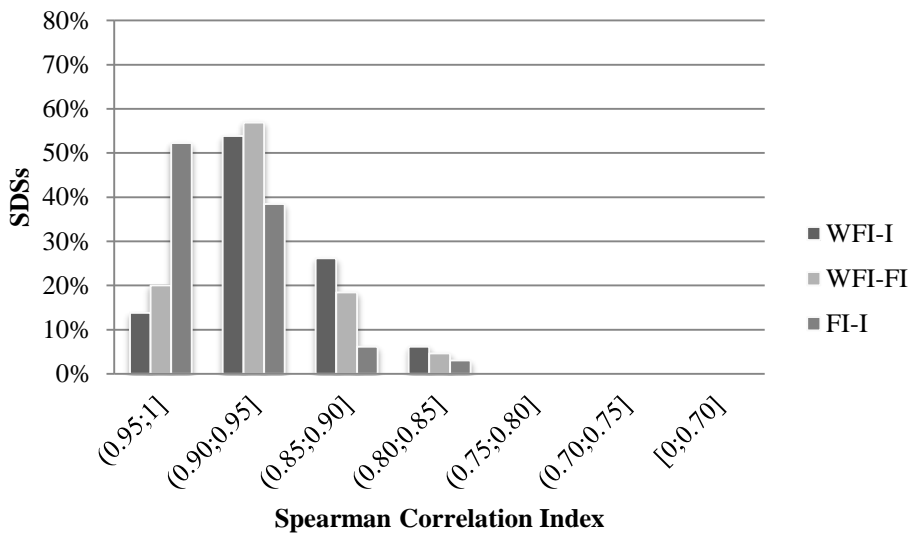

*Figure 1: Spearman correlation index for rankings based on WFI, FI and I, distribution by SDSs*

Figure 2 presents the distribution of Spearman correlation indexes for rankings based on WFO, FO and O. All the SDSs have correlations greater than 0.7. Comparing WFO and O, there are 9 SDSs (13.8% of total) with correlation greater than 0.95 and 38 SDSs (58.5%) with correlation greater than 0.9. Comparing WFO and FO, 6 SDSs (9.2% of total) show correlation between WFO and FO greater than 0.95, while for another 39 SDSs the correlation is higher than 0.9. For the comparison between FO and O there are 26 SDSs (36.9% of total) which show correlation greater than 0.95, while for 58 SDSs (89.2%) the correlation is higher than 0.9. We observe the lowest correlations for both WFO to O (0.702) and FO to O (0.680) in BIO/08 (Anthropology) and the lowest for the WFO to O comparison (0.784) in BIO/09 (Physiology).

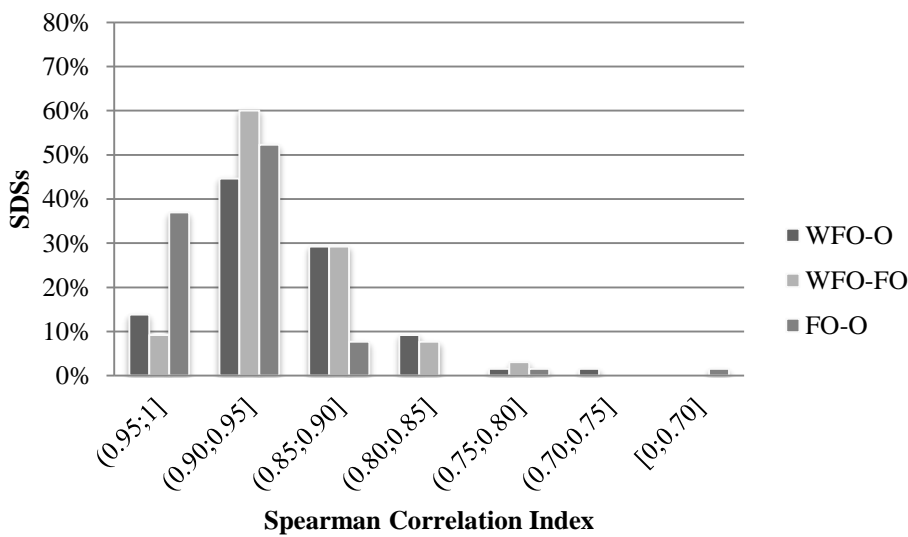

*Figure 2: Spearman correlation index for rankings based on WFO, FO and O, distribution by SDSs*

In Table 3 we present a summary of the correlation analyses obtained by comparing rankings derived from the different productivity indicators, averaging the data for the SDSs of each UDA. The spearman correlation is strongest for the rankings from FI and I (0.928 for Biology and 0.952 for Medicine) and from FO and O (0.914 for Biology and 0.944 for Medicine). The correlation values for the other ranking pairs are slightly lower.

*Table 3: Spearman correlation index for rankings based on WFI (WFO), FI (FO) and I (O), average values for SDSs of each UDA*

|           | Biology | Medicine |
|-----------|---------|----------|
| SDS       | 19      | 46       |
| WFI vs I  | 0.904   | 0.912    |
| WFI vs FI | 0.920   | 0.924    |
| FI vs I   | 0.928   | 0.952    |
| WFO vs O  | 0.885   | 0.909    |
| WFO vs FO | 0.892   | 0.906    |
| FO vs O   | 0.914   | 0.944    |

From the above analysis we see that the rankings derived from the indicators FI (FO) and I (O) show a strong correlation with the rankings by WFI and WFO. However such strong correlation does not exclude the possibility of notable shifts in rank for single universities, which could have serious consequences for the use of such ranking lists in funding decisions. With this consideration in mind we again compare the ranking lists, but this time dividing them in quartiles, analyzing the quartile shifts when we use FI (FO) or I (O) for measurement of university productivity rather than WFI and WFO. This methodology reflects the fact that categorization of organizations by quartile is very common in the national evaluation exercises intended to inform selected funding. As an example, Table 4 shows productivity quartiles for the impact indexes (column 2, 3 and 4) of the 27 universities active in MED/39 (Child Neuropsychiatry).

*Table 4: Quartiles and quartile variations of productivity measured by I, FI and WFI, for universities active in SDS MED/39 (Child neuropsychiatry)*

|            | Quartile |    |    | Quartile variations |           |         |
|------------|----------|----|----|---------------------|-----------|---------|
| University | WFI      | FI | I  | WFI vs I            | WFI vs FI | FI vs I |
| ID1        | 1        | 1  | 1  | 0                   | 0         | 0       |
| ID2        | 1        | 1  | 1  | 0                   | 0         | 0       |
| ---        |          |    |    |                     |           |         |
| ID6        | 1        | 2  | 2  | 1                   | 1         | 0       |
| ID7        | 1        | 2  | 2  | 1                   | 1         | 0       |
| ID8        | 2        | 2  | 2  | 0                   | 0         | 0       |
| ID9        | 2        | 1  | 1  | 1                   | 1         | 0       |
| ---        |          |    |    |                     |           |         |
| ID13       | 2        | 2  | 1  | 1                   | 0         | 1       |
| ID14       | 2        | 4  | 3  | 1                   | 2         | 1       |
| ID15       | 3        | 2  | 2  | 1                   | 1         | 0       |
| ID16       | 3        | 3  | 3  | 0                   | 0         | 0       |
| …          |          |    |    |                     |           |         |
| ID20       | 3        | 3  | 4  | 1                   | 0         | 1       |
| ID21       | 3        | 3  | 3  | 0                   | 0         | 0       |
| ID22       | 4        | 2  | 3  | 1                   | 2         | 1       |
| ID23       | 4        | 4  | 4  | 0                   | 0         | 0       |
| ---        |          |    |    |                     |           |         |
| ID26       | 4        | 4  | 4  | 0                   | 0         | 0       |
| ID27       | 4        | 4  | 4  | 0                   | 0         | 0       |
|            |          |    | Total | 12               | 8         | 12      |

The ID numbers in the first column are assigned according to WFI rank. For brevity, we illustrate only the examples of the first and last two universities for each quartile group under the starting WFI ranking list. The last three columns then show the absolute value of the quartile shift between rankings. Overall there are 12 quartile shifts (involving 11 out of 27 universities) between the WFI and the I rankings, 8 quartile shifts between the WFI and the FI rankings and 12 quartile shifts between the FI and the I rankings.

Table 5 shows the limit cases for the two UDAs, presenting statistics on the shifts in quartile for the two SDSs with the maximum and minimum percentages of universities registering quartile variations between WFI and I rankings. BIO/08 (Anthropology) registers the maximum percentage value of universities (47.4) affected by a quartile variation in the Biology UDA, and BIO/01 (General Botany) the minimum (16.7). In Medicine, MED/34 (Physical and Rehabilitation Medicine) registers the maximum value (50.0) and MED/17 (Infectious Diseases) shows the minimum (11.8). In similar fashion, Table 6 again shows the limit cases for quartile variations, here comparing WFO and O rankings for the two UDAs.

*Table 5: Statistics for quartile variations of productivity measured by WFI and I: for each UDA, the table shows the two SDSs with the maximum and minimum percentages of universities registering quartile variations*

| UDAs | SDS | Universities registering quartile variations (%) | Average quartile variation | Max quartile variation |
|---|---|---|---|---|
| Biology | BIO/08 | 47.4 (max value in the UDA) | 0.5 | 2 |
| | BIO/01 | 16.7 (min value in the UDA) | 0.2 | 1 |
| Medicine | MED/34 | 50.0 (max value in the UDA) | 0.5 | 1 |
| | MED/17 | 11.8 (min value in the UDA) | 0.1 | 1 |

*Table 6: Statistics for quartile variations of productivity measured by WFO and O: for each UDA, the table shows the two SDSs with the maximum and minimum percentages of universities registering quartile variations*

| UDAs | SDS | Universities registering quartile variations (%) | Average quartile variation | Max quartile variation |
|---|---|---|---|---|
| Biology | BIO/08 | 60.0 (max value in the UDA) | 0.7 | 2 |
| | BIO/02 | 17.2 (min value in the UDA) | 0.2 | 1 |
| Medicine | MED/49 | 47.1 (max value in the UDA) | 0.5 | 2 |
| | MED/40 | 12.8 (min value in the UDA) | 0.2 | 2 |

### 3.2 Correlation analysis and quartile variations in productivity rankings for UDA

In this section we apply formulas [3] and [4] as presented in part 2.1 to calculate the indicators WFI (WFO), FI (FO) and I (O) and then compare the UDA ranking lists derived from the indicators. Table 7 shows strong correlations between all the ranking lists of universities, for both UDAs, in particular between the ranking lists by FI and I (correlation index equal to 0.976 for Biology and 0.974 for Medicine) and between FO and O (correlation index of 0.912 for Biology and 0.969 for Medicine). All the other correlation values are also greater than 0.9 with the exception of those from comparing WFO and FO in Biology.

*Table 7: Spearman correlation index for rankings based on WFI (WFO), FI (FO) and I (O), by UDA*

|            | Biology | Medicine |
|------------|---------|----------|
| Universities | 61    | 54       |
| WFI vs I   | 0.934   | 0.928    |
| WFI vs FI  | 0.952   | 0.928    |
| FI vs I    | 0.976   | 0.974    |
| Universities | 63    | 54       |
| WFO vs O   | 0.923   | 0.922    |
| WFO vs FO  | 0.880   | 0.907    |
| FO vs O    | 0.912   | 0.969    |

In Table 8 we provide summary data on the shifts in quartile between the productivity rankings for the UDAS. In the case of comparing between the WFI and I lists in Biology, there are shifts in quartile for 27.9% of the universities evaluated, and in Medicine 20.4%. In the case of comparing the ranking lists for WFO e O in Biology, there are shifts in quartile for 34.9% of the universities evaluated, and in Medicine 22.2%. Both in Biology and in Medicine, the mean quartile variation is on average equal to 0.3 for the comparison between the various ranking lists (0.2 in comparing WFI and I in Medicine). There are a maximum of two universities that shift a full two or three quartiles.

*Table 8: Statistics of quartile variations of productivity measured by WFI (WFO), FI (FO) and I (O), by UDA*

|          | Universities registering quartile variations (%) | | Average quartile variation | | Universities registering quartile variations ≥2 | |
|----------|-----------|-----------|-----------|-----------|-----------|-----------|
| UDA      | WFI vs I  | WFI vs FI | WFI vs I  | WFI vs FI | WFI vs I  | WFI vs FI |
| Biology  | 27.9      | 26.2      | 0.3       | 0.3       | 1 out of 61 | 0 out of 61 |
| Medicine | 20.4      | 27.8      | 0.2       | 0.3       | 1 out of 54 | 1 out of 54 |
|          | WFO vs O  | WFO vs FO | WFO vs O  | WFO vs FO | WFO vs O  | WFO vs FO |
| Biology  | 34.9      | 30.2      | 0.3       | 0.3       | 0 out of 63 | 1 out of 63 |
| Medicine | 22.2      | 29.6      | 0.3       | 0.3       | 2 out of 54 | 2 out of 54 |

### 3.4 Analysis of top universities

For the next stage of analysis we identify the universities included in the first quartile under WFI ranking, for each SDS. Then we check which of these would not be included in the same top quartile under rankings constructed with FI and I. In a context where the funding system is conceived so as to reward only excellence, by allocating major resources to "top quartile" universities, potential shifts in ranking due to the choice of counting method would be critically important.

In Biology, an average of 24.3% of the universities considered excellent under WFI would no longer be "top" under ranking constructed with I (Table 9) and 21.8% of them would not reach top status under FI. At the SDS level (Table 10), the data for BIO/08 (Anthropology) are particularly notable: of the five universities that place in the first quartile for WFI, two (40%) are not at the top for I.

The results of the analysis for Medicine present less critical but still notable concerns: of the universities ranked as excellent under WFI, 18.6% fail to reach this status under I, and similarly 18.3% are no longer excellent under ranking constructed with FI (Table 9). We draw attention to the case of MED/01 (Medical Statistics), where three of the seven universities (42.9%) that qualify in the first quartile for WFI are not at the top for I.

*Table 9: Top 25% universities by WFI (O), not included in the same set when productivity is measured by I (O) or FI (FO)*

|     | Percentage of top 25% universities by WFI not included in the same set by | | Percentage of top 25% universities by WFO not included in the same set by | |
| --- | --- | --- | --- | --- |
| UDA | I | FI | O | FO |
| Biology | 23.4 | 21.8 | 23.7 | 22.6 |
| Medicine | 18.6 | 18.3 | 20.1 | 21.7 |

Under the output indicators, the percentages of universities that drop quartile are generally still higher, both in Biology and Medicine. In Biology, of the universities that are excellent for WFO, an average of 23.7% do not result as such in ranking constructed with O, and 22.6% fail to reach that status under FO. The case of BIO/19 (General Microbiology) is notable: of the nine universities that place in the first quartile under WFO ranking, four (44.4%) are not at the top for O.

In Medicine, 20.1% of the excellent universities under WFO are no longer top under O, and similarly 21.7% fail to reach excellent status under FO. We note the case of the MED/39 (Child neuropsychiatry) SDS: here, three of the seven universities that are in first quartile for WFO (42.9% of total) are not at the top for O.

*Table 10: Top 25% universities by WFI (WFO), not included in the same set when productivity is measured by I (O) or FI (FO): values for SDSs showing the maximum percentage in each UDA*

|     | Percentage of top 25% universities by WFI not included in the same set by | | Percentage of top 25% universities by WFO not included in the same set by | |
| --- | --- | --- | --- | --- |
| UDA | I | FI | O | FO |
| Biology | BIO/08 (40.0) | BIO/04 (33.3) | BIO/19 (44.4) | BIO/19 (44.4) |
| Medicine | MED/01 (42.9) | MED/09 (50.0) | MED/39 (42.9) | MED/39 (42.9) |

**4. Conclusions**

Evaluation at the university level often informs incentive systems and selective resource allocation. Bibliometricians are thus called on to refine techniques and indicators that can render the evaluation tools ever more accurate and robust. The shared contributions of the various researchers in co-authoring scientific advancement must certainly be taken account in individual and university evaluations. In the life sciences there are widespread and consolidated practices to signal the specific contributions of each individual author to the research outputs. Many bibliometric indicators and national evaluation exercises ignore this issue, failing to consider the order of the co-authors in the byline, and in many cases even their number. In this paper we have indicated some orders of magnitude for the distortion in university productivity rankings occurring under such circumstances.

The benchmark indicators we used to measure labor productivity are the less refined "weighted fractional output" (WFO), and the more sophisticated and accurate "weighted fractional impact" (WFI). Starting from these two productivity indicators, we measured another four: two of these, FO and FI, eliminate the weighting that accounts for the position of authors in the byline; a further two, O and I, completely ignore co-authorships.

Comparing the ranking lists for each indicator and for each field of research then permitted us to reveal the shifts from the respective benchmarks. With respect to a previous work where the authors assessed the distortions in rankings of individual researcher productivity (Abramo et al., 2012), when we move to evaluation of the entire institutions the various types of ranking lists are found to be more correlated, with slightly less but still notable distortion.

In effect, in spite of the strong general correlation between the ranking lists, the extent of shifts in university productivity rankings does not seem at all negligible. In Biology, 27.9% of universities register quartile variations between WFI and I rankings, and in Medicine the same occurs for 20.4% of the universities. In the comparison between WFO and O, a still greater share of universities experience quartile variations. Nor can we ignore the average shifts in quartile: comparing the rankings by I and WFI, the average shift is 0.3 quartiles for the universities active in Biology and 0.2 for those active in Medicine. The situation of the BIO/08 and MED/08 fields is particularly problematic, with half of the active universities placing in different quartiles if evaluated under I instead of WFI. The analysis of the institutions at the top quartile of the rankings reveals further critical concerns: in Biology, almost a quarter of the top-classed universities under WFI (first performance quartile), lose this status if evaluated under I, and in Medicine this occurs for almost 20% of top-performing institutions.

Different to the comparison of these rankings at the individual level, in the analysis at the organizational level the correlation for rankings by WFI and FI does not improve over the WFI to I correlation. This same observation can also be made in comparing the shifts in quartile under these WFI-FI and WFI-I pairings: the evidence is thus that simple fractional counting of authors, with attribution of the same share of credit to all, does not guarantee more accurate measure of productivity than measures obtained from "full" counting of authorship. This detail in the results was not particularly expected, and should be subject to further verification to determine if it might be traced to the convention used in allocating co-author contributions on the basis of their position in the byline. Another result of definite interest is the greater correlation between impact rankings compared to those for output: fractionalization of authorship seems to have less effect when applied to the citations than it does when applied to the publications themselves.

Given a context where collaboration in life sciences research is ever more the norm (98.2% of Italian Biology and Medicine articles are co-authored and 93.6% are by more than two authors), using bibliometric indicators that ignore co-authorship and real contribution reduces the accuracy and reliability of productivity measures at the university level, undermining the effectiveness of evaluation processes and compromising the results at the micro and macro-economic level.

# Appendix – List of SDSs

| UDA | SDS_code | SDS_title | UDA | SDS_code | SDS_title |
|---|---|---|---|---|---|
| | MED/01 | Medical Statistics | | BIO/01 | General Botany |
| | MED/02 | History of Medicine* | | BIO/02 | Systematic Botany |
| | MED/03 | Medical Genetics | | BIO/03 | Environmental and Applied Botany |
| | MED/04 | General Pathology | | BIO/04 | Vegetal Physiology |
| | MED/05 | Clinical Pathology | | BIO/05 | Zoology |
| | MED/06 | Medical Oncology | | BIO/06 | Comparative Anatomy and Cytology |
| | MED/07 | Microbiology and Clinical Microbiology | | BIO/07 | Ecology |
| | MED/08 | Pathological Anatomy | Biology | BIO/08 | Anthropology |
| | MED/09 | Internal Medicine | | BIO/09 | Physiology |
| | MED/10 | Respiratory Diseases | | BIO/10 | Biochemistry |
| | MED/11 | Cardiovascular Diseases | | BIO/11 | Molecular Biology |
| | MED/12 | Gastroenterology | | BIO/12 | Clinical Biochemistry and Molecular Biology |
| | MED/13 | Endocrinology | | BIO/13 | Applied Biology |
| | MED/14 | Nephrology | | BIO/14 | Pharmacology |
| | MED/15 | Blood Diseases | | BIO/15 | Pharmaceutical Biology |
| | MED/16 | Rheumatology | | BIO/16 | Human Anatomy |
| | MED/17 | Infectious Diseases | | BIO/17 | Histology |
| | MED/18 | General Surgery | | BIO/18 | Genetics |
| | MED/19 | Plastic Surgery | | BIO/19 | General Microbiology |
| | MED/20 | Pediatric and Infant Surgery | | | |
| | MED/21 | Thoracic Surgery | | | |
| | MED/22 | Vascular Surgery | | | |
| | MED/23 | Cardiac Surgery | | | |
| Medicine | MED/24 | Urology | | | |
| | MED/25 | Psychiatry | | | |
| | MED/26 | Neurology | | | |
| | MED/27 | Neurosurgery | | | |
| | MED/28 | Odonto-Stomalogical Diseases | | | |
| | MED/29 | Maxillofacial Surgery | | | |
| | MED/30 | Eye Diseases | | | |
| | MED/31 | Otorhinolaryngology | | | |
| | MED/32 | Audiology | | | |
| | MED/33 | Locomotory Diseases | | | |
| | MED/34 | Physical and Rehabilitation Medicine | | | |
| | MED/35 | Skin and Venereal Diseases | | | |
| | MED/36 | Diagnostic Imaging and Radiotherapy | | | |
| | MED/37 | Neuroradiology | | | |
| | MED/38 | General and Specialized Pediatrics | | | |
| | MED/39 | Child Neuropsychiatry | | | |
| | MED/40 | Gynecology and Obstetrics | | | |
| | MED/41 | Anesthesiology | | | |
| | MED/42 | General and Applied Hygiene | | | |
| | MED/43 | Legal Medicine* | | | |
| | MED/44 | Occupational Medicine | | | |
| | MED/45 | General, Clinical and Pediatric Nursing | | | |
| | MED/46 | Laboratory Medicine Techniques | | | |
| | MED/47 | Nursing and Midwifery* | | | |
| | MED/48 | Neuropsychiatric and Rehabilitation Nursing** | | | |
| | MED/49 | Applied Dietary Sciences | | | |
| | MED/50 | Applied Medical Sciences | | | |

*\* Excluded from the analysis since bibliometric techniques are not sufficiently robust to calculate productivity*

*\*\* Excluded from the analysis since there were only seven universities with a research staff in the five-year period.*